\documentclass{eas}
\usepackage[dvips]{graphicx}
\usepackage[margin = 1.0in]{geometry}
\usepackage{subfig}
\usepackage{amssymb}
\usepackage{epstopdf}
\usepackage{wrapfig}

\begin{document}

\title{Temporal Decomposition Studies of GRB Lightcurves} 
\author{Narayana P. Bhat}\address{University of Alabama in Huntsville, 301 Sparkman Dr. Huntsville 35899, AL USA}
%
%
\begin{abstract}
Gamma-ray bursts (GRB) are extremely energetic events and produce highly diverse light curves. Light curves are believed to be resulting from internal shocks reflecting
the activities of the GRB central engine. Hence their temporal studies can potentially lead to the understanding of the GRB central engine and its evolution. The light curve
variability time scale is an interesting parameter which most models attribute to a physical origin e.g., central engine activity, clumpy circum-burst medium, or relativistic
turbulence. We develop a statistical method to estimate the GRB minimum variability time scale (MVT) for long and short GRBs detected by GBM. We find that the MVT of short
bursts is distinctly shorter than that for long GRBs supporting the possibility of a more compact central engine of the former.  We find that MVT estimated by this method is
consistent with the shortest rise time of the fitted pulses. Hence we use the fitted pulse rise times to study the evolution of burst variability time scale. Variability 
time is in turn related to the minimum bulk Lorentz factor. Using this we relate the GRB spectral evolution to the evolution of the variability time scale.
\end{abstract}
\maketitle
\section{Introduction}
Gamma-ray bursts (GRBs) are short, intense and distant flashes of $\gamma$-rays that occur
at random locations in the sky with their peak power in the 200-500 keV range.
During their appearance, they often outshine all other sources in the $\gamma$-ray
sky combined. The temporal structure of GRBs exhibits diverse morphologies. They can vary
from a single smooth pulse to extremely complex light curves
with many erratic pulses with different durations, amplitudes,
and fine structures. Physically, several mechanisms have been invoked to interpret
GRB temporal variability. The leading scenario is to attribute the
light curve variability to the activity of the central engine (\cite{rm94,sp97}).
There are no direct observations of the Ôcentral engineÕ.
Most of the bursts exhibit variability on time scales that are much shorter than the burst durations.
According to internal shock model of GRBs, the $\gamma$-ray light curves
result from collisions between shells with different values of the bulk Lorentz factor $\Gamma$.
Within such a scenario, the observed light curves can be directly connected to the behavior of the central
engine (\cite{lw07,lz08}).

One approach for probing light curves which has received attention (\cite{nor05, bhat12}) is to express them
as a series of displaced pulses, each with a parametric form.  There is an appeal to this approach because fitting routines
are well-understood and interpretations of rise time, decay time, full width at half max, {\it etc.}, are possible.

GRB light curves exhibit variability on various time scales. 
In this paper we present a new statistical method of estimating the minimum value of such variability
time scales of a GRB and relate it to the minimum value of the fitted pulse rise time.
This in turn can be related to the minimum Lorentz factor of the relativistic shells
emitted by the central engine. The evolution of the minimum Lorentz factor is then related to
the spectral evolution of the GRBs.

\section{Minimum Variability Time-scale in GRBs}
GRB light curves are generally binned in to narrow time bins. Such light curves with high 
variability at low power may show variations
which are not statistically significant. While statistically significant variability could become 
statistically insignificant at finer bin-widths and significant variability could vanish if the
bin-widths are too coarse. By a comparison of the GRB prompt emission variability with the
purely statistical variability of the background region we derive an optimum bin-width when the
non-statistical variability in the light curve becomes significant.

\begin{figure}
\centerline{\includegraphics[width=8.cm]{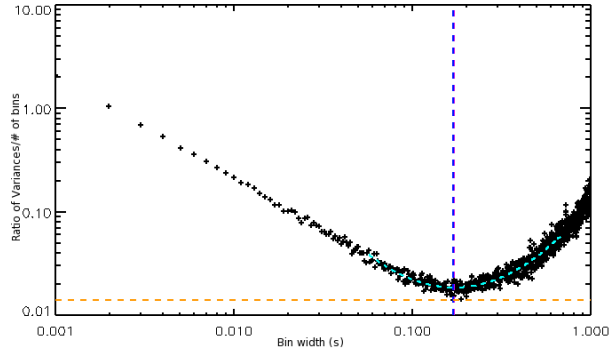}}
\caption{Variation of the ratio of the variances per bin to the histogram bin-width. 
At very fine bin-widths the GRB signal is indistinguishable from background
fluctuations and hence the ratio decreases monotonically with increasing bin-width.
At larger bin-widths the signal is clearly visible from the background and hence the 
ratio per bin starts increasing. The turn over bin-width is defined as the minimum
variability time scale where the bin-width is optimum. Cyan dashed line shows a fitted parabola around the minimum that has a minimum
at a bin-width indicated by the vertical dashed line in blue. See text for more details.
\label{var_ts}}
\end{figure}

We identify the prompt emission duration and an equal duration of background region. We then derive a
differential of each light curve and compute the ratio of the variances of the GRB and the background.
This ratio divided by the bin-width is plotted as a function of bin-width in figure \ref{var_ts}.
As can be seen at very fine bin-widths the ratio falls monotonically with increasing bin-width
signifying that at such fine bin-widths the variations in the background and burst regions are statistically identical.
In other words the signal in the burst light curve is indistinguishable from Poissonian fluctuations.
Later at certain bin-width the variation starts deviating from the $\frac {1} {bin-width}$ behavior. We measure the
bin-width at this valley by fitting a parabola and the bin-width at the minimum of the parabola is called the optimum bin-width $t_b$.
$t_b$ is also interpreted as the minimum variability time scale of the GRB $t_v$. At this bin-width the variability in the GRB light curve
becomes detectable compared to the normal fluctuations seen in the background. We are therefore
confident that the short-term variability is real and is not an
artifact of data reduction or statistical fluctuations.

Another possibility is that $t_b$ could be a function of the signal-to-noise ratio of the GRB rather than an intrinsic feature of the GRB light curve.
For this we generated synthetic light curves of the same GRB by adding Poisson noise to the fitted lognormal pulses and adding them to the fitted background
with added noise. The optimum bin-width was estimated for each of the synthetic light curve derived by varying the signal-to-noise ratio over a few orders of
magnitude. It was found that the value of $t_b$ is not a strong function of the signal-to-noise ratio.

\begin{figure}
\centerline{\includegraphics[width=8.cm]{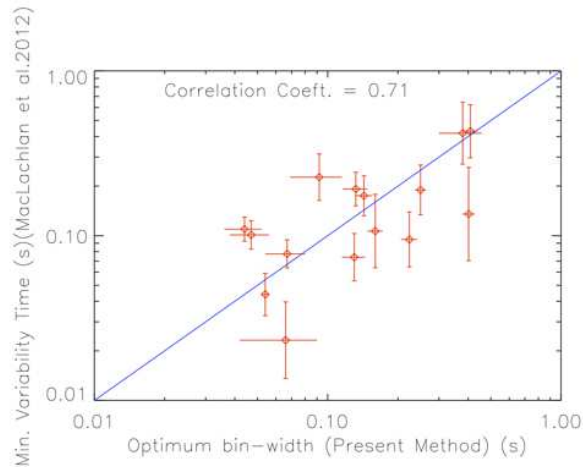}}
\caption{A comparison of the minimum variability time scale estimated by McLachlan et al.(2012) with the
optimum bin-width estimated by the present method. They seem to correlate well.
\label{ts_ts}}
\end{figure}

Recently the minimum variability time scales were estimated for long and short GRBs by a model independent 
method based on a wavelet decomposition technique (\cite{mac12}). In figure \ref{ts_ts} we compare the minimum variability time scales estimated by MacLachlan \etal\ (2012) 
with those estimated by the present method for the same GRBs. The blue line shows the ideal case when the two quantities are equal.
The figure shows that the minimum variability time scales estimated by two different techniques are statistically consistent with each other. Hence we conclude that
the present method indeed estimates the minimum variability time scale of a given GRB.  MacLachlan \etal\ (2012) also demonstrate that the minimum variability time scale
estimated by the wavelet decomposition technique is also consistent with the minimum of the rise-times of the fitted pulses using the lognormal shape 
for the individual pulses (\cite{bhat12}). 

\subsection{Minimum Variability Time-scale as a GRB type identifier}

\begin{figure}
\centerline{\includegraphics[width=8.cm]{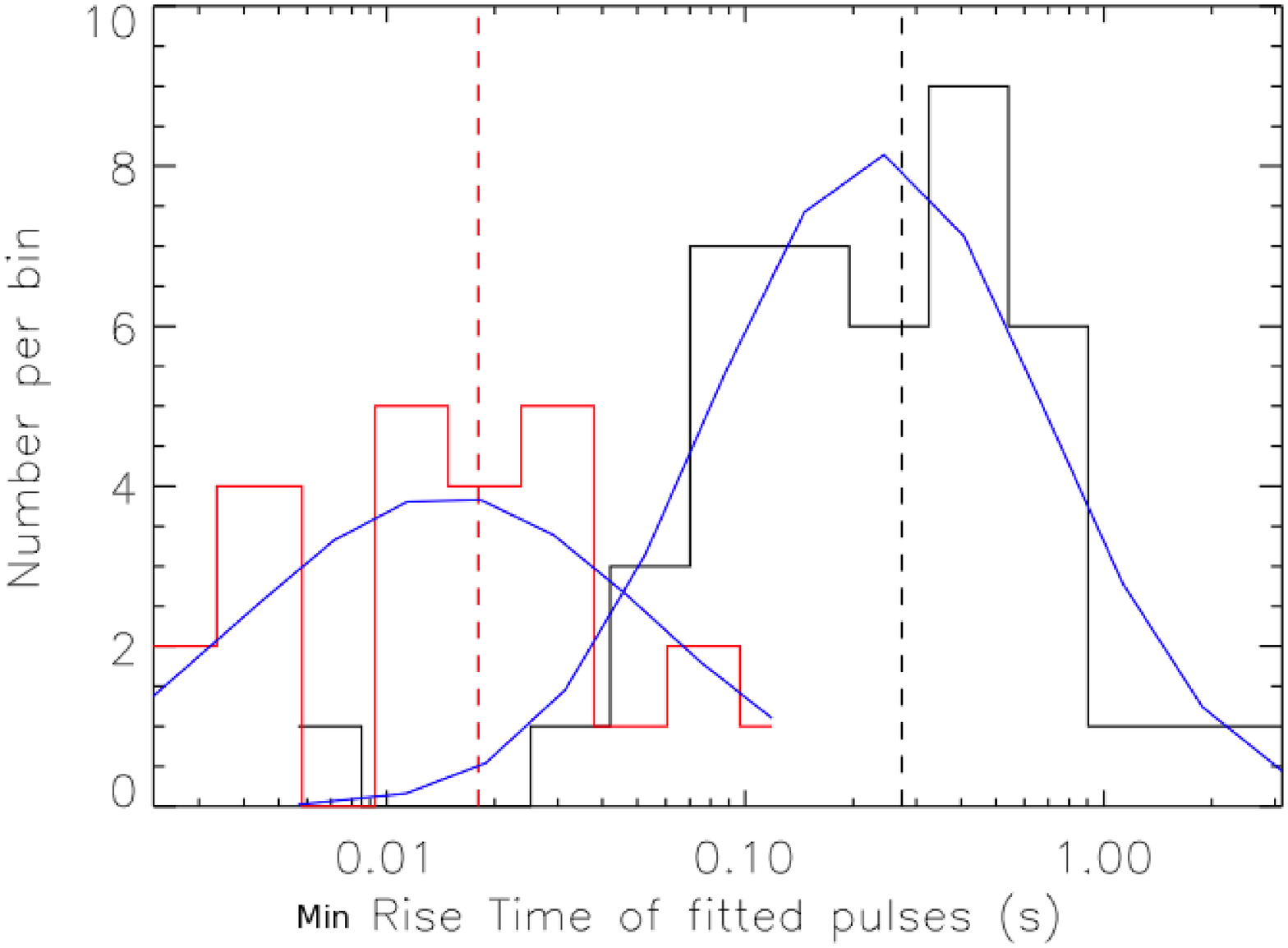}}
\caption{A distribution of the minimum pulse rise time for long and short GRBs.
The minimum variability time scale or the minimum fitted pulse rise time clearly shows a 
bimodal distribution showing that it can be a parameter to distinguish between long and
short GRBs.
\label{mts_dist}}
\end{figure}

GRBs are generally classified as long or short depending on their duration $T_{90}$ and their hardness ratios.
Long GRBs ($T_{90}\,>\,2s$) are generally soft and short GRBs ($T_{90}\,<\,2s$) are generally hard. They are considered to form two different GRB classes (\cite{kmf93}).
The measurement of spectral lags is another tool in the study
of GRBs and their classification since short GRBs exhibit negligible lags. However classification schemes based on any or all of these parameters result in significant
overlap of GRBs of either type. Hence we need more such identifying parameters to uniquely identify a GRB type. Here we have another parameter, $t_v$, to add to that list.
Figure \ref{mts_dist} shows a distribution of minimum rise-times for long and short GRBs. The minimum pulse rise-times
of short GRBs are distinctly shorter (at least by a factor of 15) than that of long GRBs. In other words the minimum variability time scale, $t_v$, can be used as 
another parameter to identify short GRBs.

\section{Minimum Variability Time scale and minimum Lorentz factor}

\begin{figure}
  \centering
  \subfloat[An example of evolution of the pulse width (FWHM) as a function of energy as seen in GRB1107311A. ]{\label{f1}\includegraphics[width=0.42\textwidth]{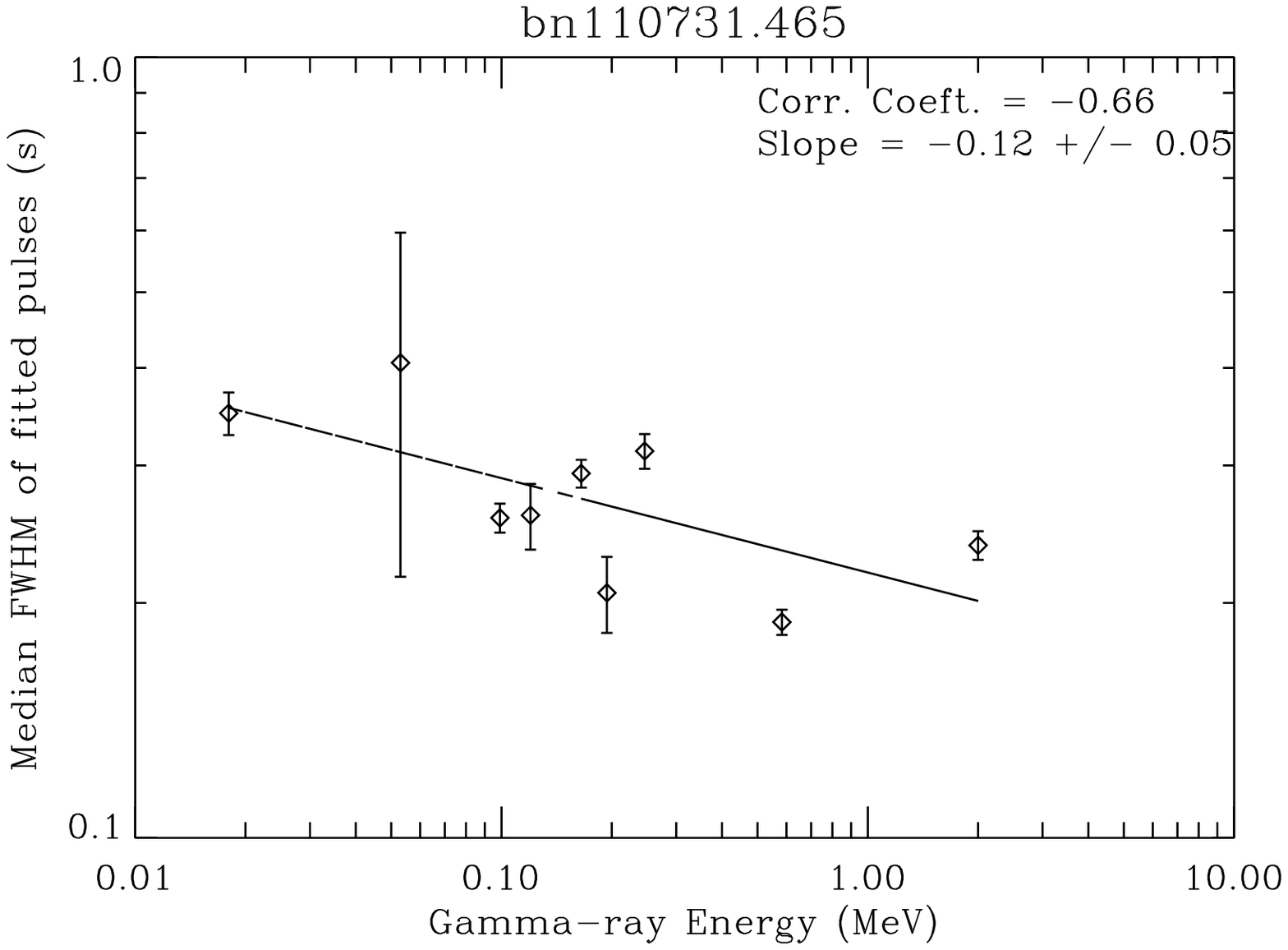}}
  \hspace{1.0cm}
  \subfloat[Evolution of the pulse width (FWHM) as a function of time since trigger as seen in GRB1107311A.]{\label{f2}\includegraphics[width=0.42\textwidth]{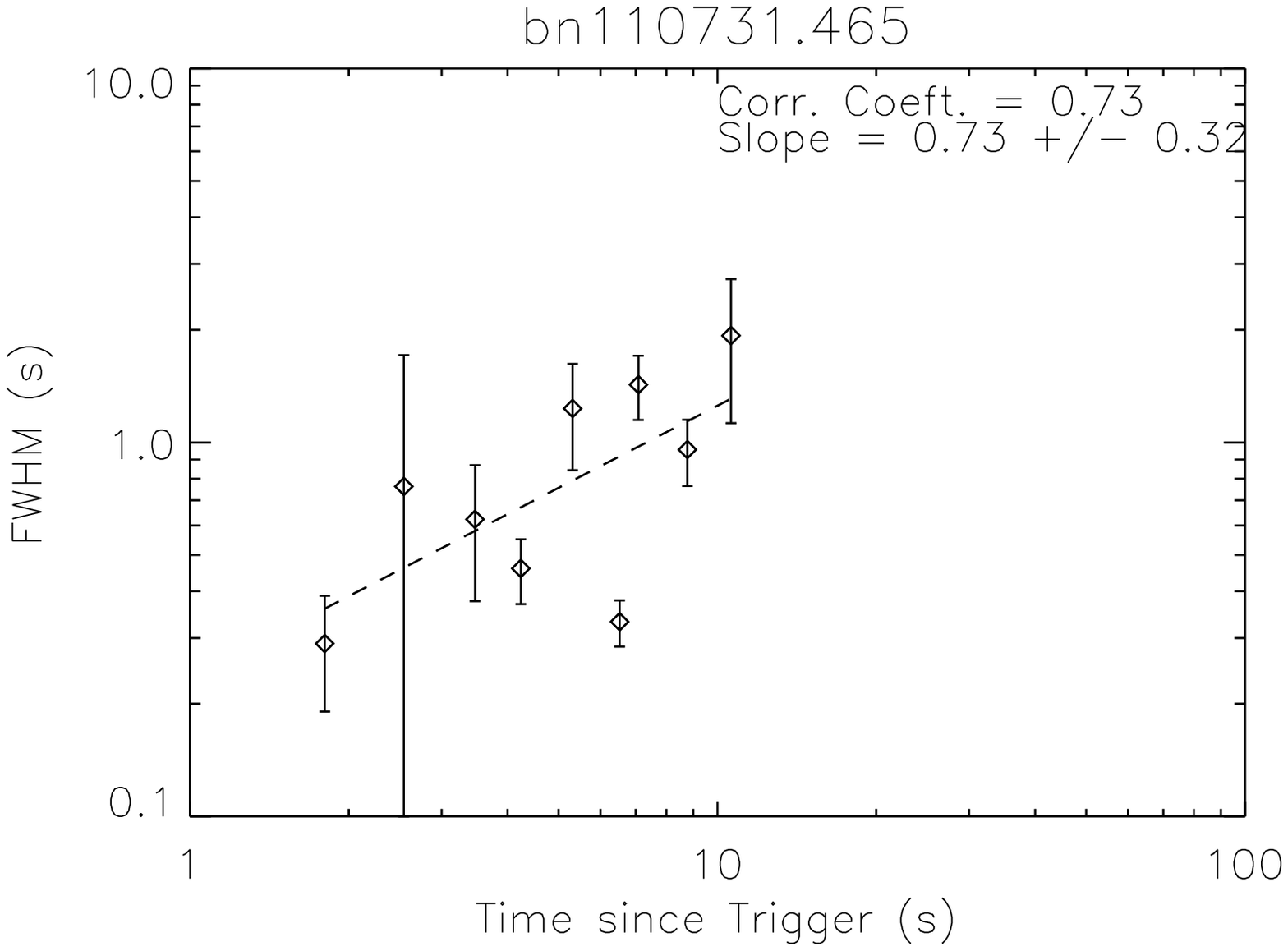}}
\caption{}
\label{fig12}
\end{figure}

\begin{figure}
\centerline{\includegraphics[width=8.cm]{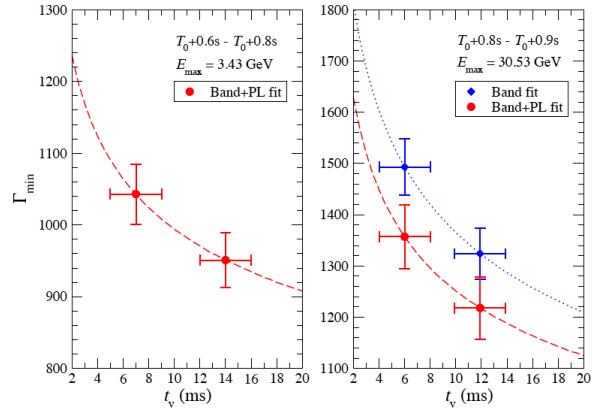}}
\caption{The $\Gamma_{\rm min}$ of the GRB 090510 prompt mission region as a function of the $\gamma$-ray variability time-scale $t_v$. The highest and low-energy (MeV)
photons in each time-interval are assumed to originate from the same physical region. The data points correspond to the $\Gamma_{\rm min}$ calculated for the best-fit
$t_v =$ FWHM and FWHM/2 in the respective time intervals.
\label{gam_ev}}
\end{figure}

While $t_v$ has been shown to be consistent with the minimum rise-time of fitted pulses, the rise-times of the fitted pulses to the prompt emission light curve
can be used to trace the evolution of the variability time scale at any time during the GRB prompt emission.
One can also derive a lower limit on the bulk
Lorentz factor ($\Gamma_{min}$) given the variability timescales and observations of the highest energy photons
at any time during a burst. Constraining the bulk Lorentz factor, $\Gamma$, of the jet is a major
challenge in understanding the GRB physics, the mechanism for launching the jet as well
as high-energy emission. High-energy $\gamma$-rays produced and emitted from the GRB jet are
subject to $\gamma\gamma~\rightarrow~e^+e^-$ pair production process with soft target photons, and absorbed {\it in\,situ}.
The interaction rate of this process and corresponding opacity, $\tau_{\gamma\gamma}$, for the high-energy $\gamma$-rays
depends on the target photon density and can be significant when both the high-energy and
target photons are produced in the same physical region. Highly relativistic motion, with
a bulk Lorentz factor $\Gamma\gg 1$, of such an emission region can reduce the $\gamma\gamma$ interaction rate
and $\tau_{\gamma\gamma}$ greatly by allowing for a larger emitting radius and a smaller target photon density.
Observation of a $\gamma$-ray spectrum up to an energy $E_{max}$ thus can be used to put a lower limit
on $\Gamma$ (\cite{lith01, razz04,gran08,acke10}).
Thus the evolution of the variability time scale leads to the study of the evolution of the bulk Lorentz factor which in turn is related to the evolution of the $\gamma$-ray opacity $\tau_{\gamma\gamma}$ during the
prompt emission phase of a GRB. A delayed onset of the GeV photons, seen in several GRBs detected in the Fermi LAT, the emission is interpreted as due to the time evolution of the opacity in a GRB outflow (\cite{has12}).  In addition, As pointed out by Granot et al. (2008), due to the temporal evolution of $\tau_{\gamma\gamma}$, the opacity cut-off in a time-integrated spectrum will be
smoother than a sharp exponential decay: the cut-off transition will
be close to a power-law steepening. This
time evolution takes place within a given $\gamma$-ray pulse, and can
be even stronger in a complex burst where the light curve is made
up of many pulses (\cite{aoi10}). Hence a study of the evolution of $\Gamma_{min}$ would lead to a better understanding of the possible connection between the temporal structure of the light curve and the spectral evolution of the GRB.  Figure \ref{fig12}a shows variation of the fitted pulse width (FWHM) as a function of $\gamma$-ray energy while figure \ref{fig12}b shows a
variation of the FWHM as a function of time since the GRB trigger time. It is well known that for lognormal pulse shapes the rise-time and FWHM are strongly correlated
(\cite{bhat12}). Hence figure \ref{fig12}a demonstrates that the variability time scale decreases with increasing $\gamma$-ray energy while figure \ref{fig12}b shows that the 
variability time scale decreases since the trigger, indicating a trend like the hard-to-soft spectral evolution in GRBs. Figure \ref{gam_ev} shows a typical example of the 
observed $\Gamma _{min}$ evolution during a bright short Fermi GRB 090510(\cite{abdo09}). Here the $\Gamma _{min}$ is estimated at two different epochs of the GRB (0.6\,s\,-\,0.8\,s, left panel and 0.8\,s\,-\,0.9\,s, right panel, post trigger) by assuming that the
highest energy of the $\gamma$-ray ($E_{max}$) emitted in each interval originate from the same physical region as the observed low energy photons in the same time interval. The $t_v$ were approximated conservatively to the FWHM of the fitted pulse during each interval. The data points with error bars correspond to the  $\Gamma _{min}$ calculated for the best-fit  $t_v$ = FWHM and FWHM/2 in the respective time intervals.

\section{Summary}
A method is developed to estimate the optimum bin-width of the light curve to carry out pulse decomposition analysis of GRBs. The optimum bin-width is interpreted
as the minimum variability time scale of the GRB because it is found to be statistically consistent with that estimated by an independent method which in turn
is found to be consistent with the lowest of the rise-times of the fitted pulses to deconvolve the entire light curve of the GRB. 
The variability time scales were also estimated using the {\it fwhm} of the fitted pulses during the course of the burst and study the evolution of the bulk Lorentz factor.
The variability time scales, $t_b$, were estimated at different $\gamma$-ray energies by the same pulse fitting technique using GRB light curves in different energy 
ranges. Using the estimated bulk Lorentz factor one can test the location of $\gamma$-ray emission regions in the internal shock scenario. The $\gamma$-ray emission radius 
is given by $R\sim 2\Gamma^2ct_b/(1+z)$. Thus an energy dependent variation of $t_b$ can be interpreted as collisions of shells at different radii producing $\gamma$-rays of
different energies.


\end{document}